\documentstyle[eqsecnum,aps,epsfig]{revtex}
\begin{document}
\title{GAUGE DEPENDENCE OF MASS AND CONDENSATE IN CHIRALLY
ASYMMETRIC PHASE OF QUENCHED QED3}
\author{A. Bashir, A. Huet and A. Raya} 
\address{Instituto de F{\'\i}sica y Matem\'aticas, 
         Universidad Michoacana de San Nicol\'as de Hidalgo\\
         Apartado Postal 2-82, Morelia, Michoac\'an 58040, M\'exico.}
\maketitle
\begin{abstract} 

We study three dimensional quenched Quantum Electrodynamics in the bare 
vertex approximation.
We investigate the gauge dependence of the dynamically generated Euclidean
mass of the fermion and the chiral condensate for a wide range of values
of the covariant gauge parameter $\xi$. We find that 
(i) away from $\xi=0$, gauge dependence of the said quantities
is considerably reduced without resorting to sophisticated vertex 
{\em ansatze}, (ii) wavefunction renormalization plays an
important role in restoring gauge invariance and (iii) the 
Ward-Green-Takahashi identity seems to increase the gauge dependence when 
used in conjunction 
with some simplifying assumptions. In the Landau gauge, we also verify 
that our results are in agreement with those based 
upon dimensional regularization scheme within the numerical accuracy 
available. \\

\end{abstract}

\hspace{16mm}PACS number(s): 11.10.Kk,11.15.Tk,12.20.-m \hspace{35mm} UMSNH-PHYS/02-1

\newcommand{\nn}{\nonumber}
\newcommand{\kp}{k \cdot p}
\newcommand{\qp}{q \cdot p}
\newcommand{\qk}{q \cdot k}
\newcommand{\be}{\begin{eqnarray}}
\newcommand{\ee}{\end{eqnarray}}
\newcommand{\lna}{ {\rm ln}\left| \frac{k+p}{k-p} \, \right| }
\newcommand{\lnb}{\frac{1}{kp}{\rm ln}\left| \frac{k+p}{k-p} \, \right| }
\newcommand{\lnc}{\frac{k^2+p^2}{2kp}{\rm ln}\left| \frac{k+p}{k-p} \, 
\right| }
\newcommand{\lnd}{\frac{(k^2-p^2)^2}{2kp}{\rm ln}\left| \frac{k+p}{k-p} \, 
\right|}

\section{Introduction}

 In a gauge field theory, Green functions transform in a specific 
manner under a variation of gauge. These transformations carry the
name Landau-Khalatnikov-Fradkin (LKF) transformations, \cite{LK1,LK2,F1}.
These were also derived  by Johnson and Zumino through functional
methods, \cite{JZ1,Z1}. As a consequence of gauge covariance, Green
functions obey certain identities which relate one function to
the other. These relations have been named Ward-Green-Takahashi identities
(WGTI), \cite{W1,G1,T1}. At the level of physical observables, gauge symmetry
reflects as the fact that they be independent of
the gauge parameter. Perturbation theory respects these requirements at
every level of approximation. However, this has not been achieved in general
in the non-perturbative study of gauge field theories through 
Schwinger-Dyson equations (SDE) carried out so far although significant 
progress has been
made. The gauge technique of Salam, Delbourgo 
and later collaborators, \cite{Salam1,SD1,S1,DW1,DW2,D1}, was developed to 
incorporate the constraint imposed by WGTI. However, as pointed out in
\cite{D2}, gauge technique can become completely reliable only after
incorporating transverse Green functions with correct analytic and 
gauge-covariance properties. Another method widely used to explore
the non-perturbative structure of the SDE is
to make an {\em ansatz}  for the full fermion-boson vertex and then study the 
gauge dependence of the physical observables related to the phenomenon 
of dynamical chiral symmetry breaking (DCSB). This method has been quite 
popular in four dimensional Quantum Electrodynamics (QED). For 
example, the vertex {\em ansatz} proposed by Curtis and Pennington,
\cite{CP1}, has been extensively used to study the gauge dependence of the 
fermion  propagator and the dynamical generation of fermion mass in 
Quenched QED, e.g., \cite{CP2,CP3,ABGPR1,AGM1}. Later on, the work of 
Bashir and Pennington, \cite{BP1,BP2}, proposed an improved vertex which
achieves complete gauge independence of the critical coupling above which
mass is dynamically generated. These methods use the cut-off regularization 
to study the gauge dependence of the physical observables. As the cut-off
method in general does not respect gauge symmetry, a criticism of these 
works has been raised recently, \cite{GSSW,SSW,KSW}. They suggest dimensional
regularization scheme to study the chirally asymmetric phase of QED so that
the possible gauge dependence coming from the inappropriate regulator
could be filtered out.

  Three dimensional Quantum Electrodynamics (QED3) provides us with a neat 
laboratory to study DCSB as it is ultraviolet 
well-behaved and hence the source
of gauge non-invariance finds its roots only in the simplifying
assumptions employed and {\em not} in the choice of the regulator.
Burden and Roberts, \cite{BR1}, studied the gauge dependence of the
chiral condensate in quenched QED3 and proposed a vertex which 
appreciably reduces this gauge dependence in the range $0-1$ of the
covariant gauge parameter $\xi$. Unfortunately, the choice of their 
vertex does not transform correctly under the operation of charge 
conjugation. Moreover, the selected range of values for $\xi$ is very
narrow, close to the vicinity of the Landau gauge. The studies of QED3
carried out by Bashir {\em et. al.}, \cite{BKP1,AB1}, reveal that Landau 
gauge may not be a preferred gauge in QED3. In fact, it is seen 
that the LKF transformation of one loop fermion propagator remains 
insensitive to the inclusion of the LKF transformation of the second
loop, provided we are away from the Landau gauge. Assuming similar
behaviour for the higher loops, we may expect that for larger
value of the gauge parameter, physical observables would also 
become insensitive to the value of this parameter. In this paper, we 
undertake the calculation of the Euclidean mass of the fermion (referred
to as {\em mass} from now onwards) 
and the condensate for a wide range of values of $\xi$. We find that
indeed for larger values of the gauge parameter, both the quantities
seem to become increasingly gauge invariant.

\section{The Fermion Propagator}

In quenched QED3, the SDE for the fermion propagator in the Minkowski space
can be written as~:
\be
S_F^{-1}(p) &=& S_F^{0 -1}(p) - 
ie^2 \int \frac{d^3k}{(2\pi)^3} \Gamma^{\nu}(k,p) 
S_F(k) \gamma^{\mu} \Delta^0_{\mu \nu}(q) \label{prop} \;,
\ee
where $q=k-p$, $e$ is the electromagnetic coupling, $\Gamma^{\nu}(k,p)$ is
the full fermion-photon vertex, $S_F^{0}(p)$ and
$\Delta^0_{\mu \nu}(q)$ are the bare fermion and photon propagators defined
as
\be
  S_F^{0}(p)=1/\not \! p\;, \hspace{10mm} \Delta^0_{\mu \nu}(q) = 
-\frac{g_{\mu \nu}}{q^2} + (1-\xi)\frac{ q_{\mu}q_{\nu} }{q^4} 
\label{bareprop} \;,
\ee
and $S_F(p)$ is the full fermion propagator, which we prefer to write in
the following most general form~:
\be
S_F(p)=\frac{F(p)}{\not \! p- {\cal M}(p)} \label{fullprop} \;.
\ee
$F(p)$ is referred to as the wavefunction renormalization and ${\cal M}(p)$
as the mass function. $\xi$ is the usual covariant gauge parameter. 
Eq.~(\ref{prop}) is a matrix equation. It consists of
two independent equations, which can be decoupled by taking its trace after
multiplying it with $1$ and $\not \! p$, respectively. Making use of 
Eqs.~(\ref{bareprop},\ref{fullprop}) and replacing the full vertex 
by its bare counterpart, these equations can be written as~:
\begin{eqnarray}
\frac{1}{F(p)}&=& 1 + \frac{\alpha}{2 \pi^2 p^2} \int d^3k
\, \frac{F(k)}{k^2+{\cal M}^2(k)} \, \frac{1}{q^4} \;
\left[\,-2 (k \cdot p)^2 + (2-\xi) (k^2+p^2) k \cdot p - 2 (1-\xi) k^2 p^2\,
\right] \;,\\
\frac{{\cal M}(p)}{F(p)}&=&   \frac{\alpha(2+\xi)}{2 \pi^2} \int d^3k
\, \frac{F(k)\,{\cal M}(k)}{k^2+{\cal M}^2(k)} \, \frac{1}{q^2} \;.
\end{eqnarray}
Carrying out angular integration after the Wick rotation to the Euclidean 
space, the
above equations acquire the form~:
\be
\frac{1}{F(p)} &=& 1 - \frac{\alpha \xi}{\pi p^2} \int \, dk \,
\frac{k^2 F(k)}{k^2+ {\cal M}^2(k)}
\left[\; 1 - \frac{k^2+p^2}{2kp} \, {\rm ln}\left| \frac{k+p}{k-p} \, \right|
\;  \right]  \;, \label{FNoWT} \\  \nn  \\ 
\frac{{\cal M}(p)}{F(p)} &=& \frac{\alpha( \xi +2 )}{\pi p}  \int \, dk \,
\frac{k F(k)  {\cal M}(k) }{k^2+ {\cal M}^2(k)} \, 
{\rm ln}\left| \frac{k+p}{k-p} \, \right|  
\; .     \label{MNoWT} 
\ee
A trivial solution to Eq.~(\ref{MNoWT}) is ${\cal M}(p)=0$, which corresponds
to the usual perturbative solution. We are interested in a non-trivial
solution by solving Eqs.~(\ref{FNoWT}) and (\ref{MNoWT}) simultaneously.
Such a solution for ${\cal M}(p)$ is related to the
mass $m$ and the chiral condensate $\, <\bar{\psi} \psi> \,$ 
through the relations
 $m={\cal M}(m)$ and $<\bar{\psi} \psi>=4 p^2 {\cal M}(p)/(2+\xi)$, 
respectively. We shall study the gauge dependence of these quantities 
in the next section for the fixed value of $\alpha=1/4 \pi$.

\section{Effect of the Wavefunction Renormalization}

\begin{center}
\begin{figure}[h]     
\rotatebox{-90}                     
{\resizebox{10cm}{13cm}{\includegraphics{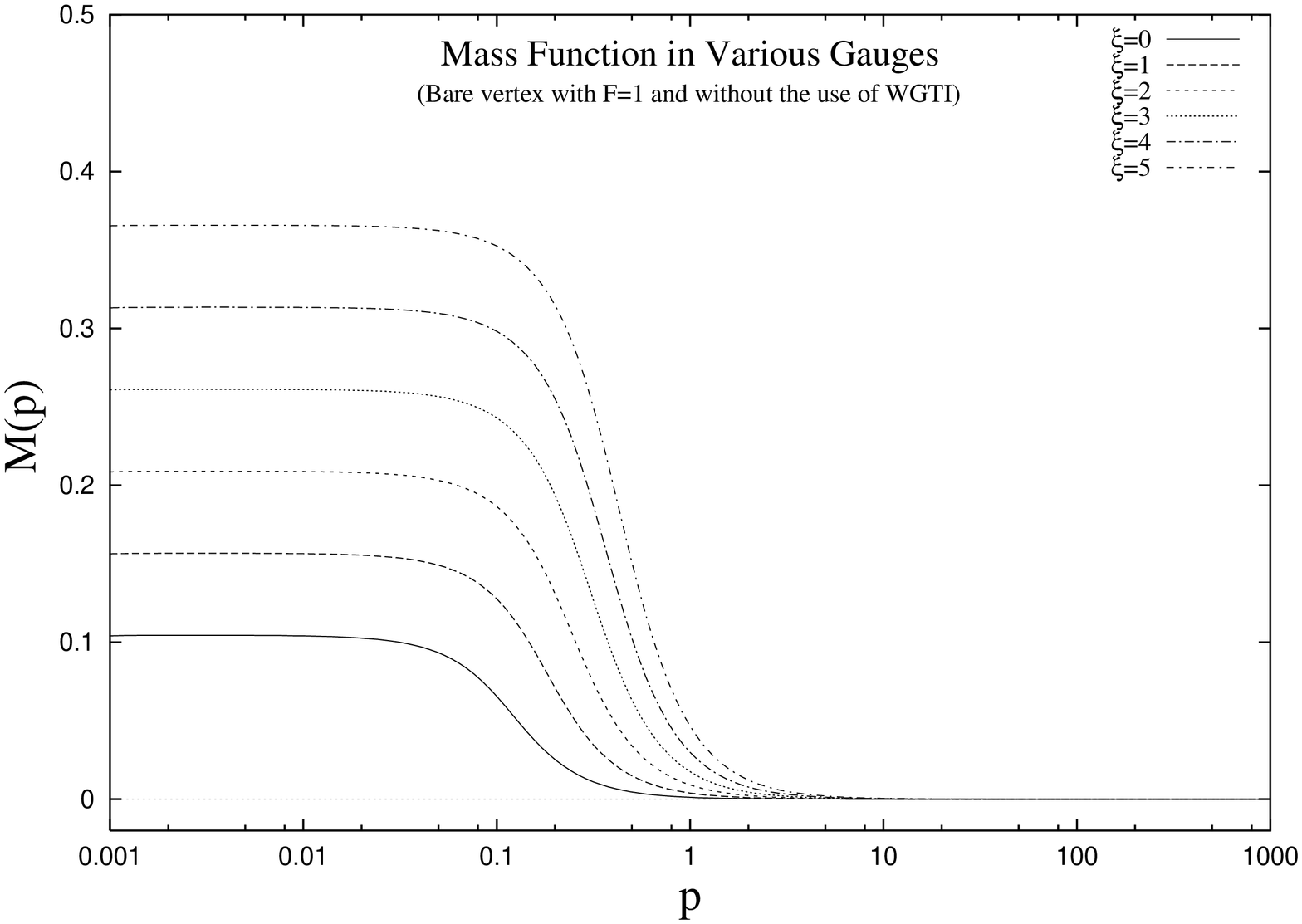}}}  
\vspace{5mm}                 
\caption{The mass function ${\cal M}(p)$ in the approximation $F(p)=1$.} 
\label{fnoWTIF1} 
\end{figure}          
\end{center}

\begin{center}
\begin{figure}[h]     
\rotatebox{-90}                     
{\resizebox{10cm}{13cm}{\includegraphics{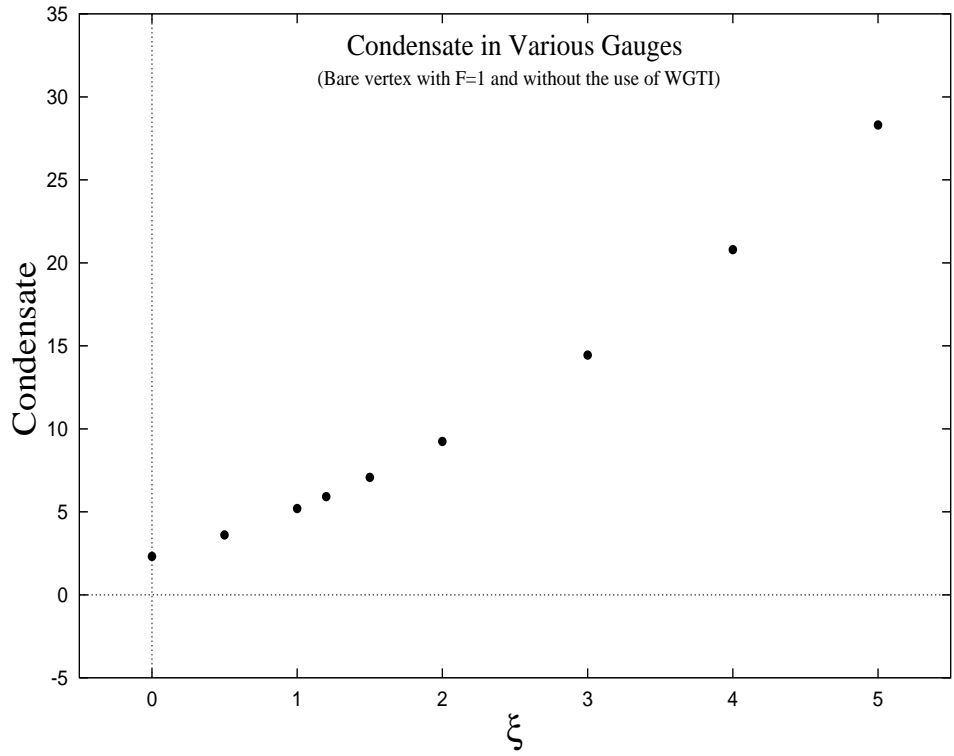}}}  
\vspace{5mm}                 
\caption{The condensate $<\bar{\psi} {\psi}>$ in the approximation $F(p)=1$.} 
\label{fnoWTIF1condensate1}                                        
\end{figure}          
\end{center}

\begin{center}
\begin{figure}[h]     
\rotatebox{-90}                     
{\resizebox{10cm}{13cm}{\includegraphics{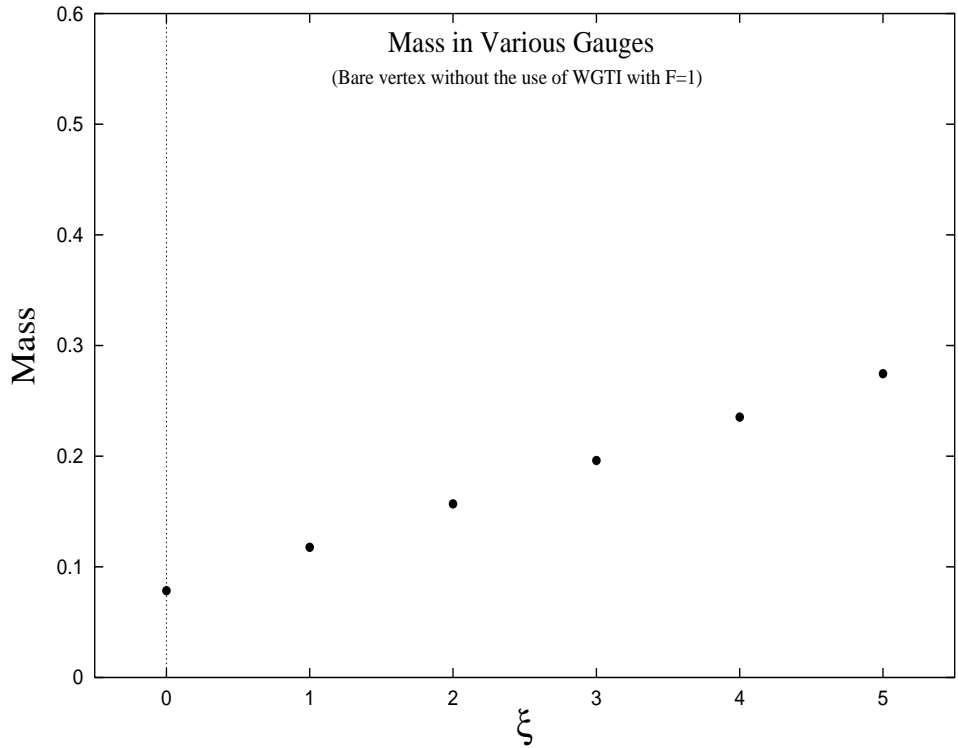}}}  
\vspace{5mm}                 
\caption{The mass in the approximation $F(p)=1$.} 
\label{fmassnoWTIF1} 
\end{figure}          
\end{center}

   In studying DCSB, it has been a common practice to make the 
approximation $F(p)=1$ so that
we only have to solve Eq.~(\ref{MNoWT}). The justification for this
approximation stems from the fact that perturbatively
$F(p)=1+{\cal O}(\alpha \xi/\pi)$. If $\alpha$ is small and we are
sufficiently close to the Landau gauge, one would naturally expect
that $F(p) \approx 1$. Although, it has been quite customary to employ this 
approximation, there exist several works which include 
both the equations. We study the effects of neglecting the wavefunction
renormalization quantitatively. Fig.~(\ref{fnoWTIF1}) depicts the mass
function ${\cal M}(p)$ for $F(p)=1$ in various gauges. As expected,
the mass function is roughly a constant for low values of $p$ and 
falls as $1/p^2$ for large values of $p$. The integration region chosen is
from $10^{-3}$ to $10^3$ and we select 26 points per decade. 
The mass 
probes low momentum region of this graph, whereas, 
the condensate is extracted from its asymptotic behaviour. 
Obviously, the mass seems to vary in more or less 
equally spaced steps with the variation of the gauge parameter.
In order to obtain a quantitative value of the mass, we select
neighbouring points $p_a$ and $p_b$ ($p_a>p_b$), such that
${\cal M}(p_a)<p_a$ and ${\cal M}(p_b)>p_b$. We 
then approximate the mass by the following relation~:
\be
m &=& \frac{{\cal M}(p_b)-{\cal M}(p_a)}{p_b-p_a}(m-p_a)+{\cal M}(p_a) \;.
\ee
 As for the 
condensate, the figure does not distinguish between the results for various 
gauges. Therefore, we have to look at the numbers explicitly. Table~(1) 
shows the value of the condensate for $\xi$ ranging from $1-5$. The point
$p=1000$ was chosen to calculate the condensate. This number seems 
sufficiently large as the $1/p^2$ behaviour seems to set in much earlier
($p \approx 300$), as noted also in \cite{BR1}. 
In Figs.~(\ref{fnoWTIF1condensate1}) and 
(\ref{fmassnoWTIF1}) we display the gauge dependence of the
chiral condensate and the mass for $F(p)=1$ in a wide range
of values of the gauge parameter. The condensate varies heavily with the 
change of gauge, roughly twice per unit change in the value of $\xi$. 
Gauge dependence of the mass is not too different either.     

\vspace{5mm}
\hspace{4.5cm
\begin{tabular}{|crr|}  \hline 
&     &    \\
\multicolumn{1}{|c}{$\xi$}  &
\multicolumn{1}{c}{$p$} &
\multicolumn{1}{c|}{$\frac{4}{2+\xi}\,p^2\,M(p^2)$}  \\
&     &    \\                                 \hline  \hline
&     &    \\
0.0 &  1000      &  2.31109     \\   
    &  642.233   &  2.31117     \\   
    &  316.228   &  2.31119     \\   
&     &    \\                        \hline
&     &    \\
0.5 &  1000      &  3.61103     \\   
    &  642.233   &  3.61119     \\   
    &  316.228   &  3.61124     \\   
&     &    \\                        \hline
&     &    \\
1.0 &  1000      &  5.19982     \\   
    &  642.233   &  5.20009     \\   
    &  316.228   &  5.20017     \\   
&     &    \\                        \hline
&     &    \\
1.2 &  1000      &  5.91622     \\   
    &  642.233   &  5.91654     \\   
    &  316.228   &  5.91664     \\   
&     &    \\                        \hline
&     &    \\
1.5 &  1000      &  7.07745     \\   
    &  642.233   &  7.07787     \\   
    &  316.228   &  7.07800     \\   
&     &    \\                        \hline
&     &    \\
2.0 &  1000      &  9.24390     \\   
    &  642.233   &  9.24452     \\   
    &  316.228   &  9.24473     \\   
  &     &    \\                               \hline   
\end{tabular}
\begin{tabular}{|crr|}  \hline 
&     &    \\
\multicolumn{1}{|c}{$\xi$}  &
\multicolumn{1}{c}{$p$} &
\multicolumn{1}{c|}{$\frac{4}{2+\xi}\,p^2\,M(p^2)$}  \\
&     &    \\                                 \hline  \hline
&     &    \\
2.5 &  1000      &  11.6992     \\   
    &  642.233   &  11.7001     \\   
    &  316.228   &  11.7003     \\   
&     &    \\                        \hline
&     &    \\
3.0 &  1000      &  14.4432     \\   
    &  642.233   &  14.4444     \\   
    &  316.228   &  14.4448     \\   
&     &    \\                        \hline
&     &    \\
3.5 &  1000      &  17.4761     \\   
    &  642.233   &  17.4777     \\   
    &  316.228   &  17.4782     \\   
&     &    \\                        \hline
&     &    \\
4.0 &  1000      &  20.7977     \\   
    &  642.233   &  20.7998     \\   
    &  316.228   &  20.8005     \\   
&     &    \\                        \hline
&     &    \\
4.5 &  1000      &  24.4081     \\  
    &  642.233   &  24.4108     \\  
    &  316.228   &  24.4117     \\   
&     &    \\                        \hline
&     &    \\
5.0 &  1000      &  28.3073     \\  
    &  642.233   &  28.3106     \\   
    &  316.228   &  28.3117     \\  
  &     &    \\                               \hline   
\end{tabular}}  \label{table1} \\ \\ \\
{\centerline {TABLE~1. The condensate in various gauges for $F(p)=1$}}
\vspace{5mm}

\begin{center}
\begin{figure}[h]     
\rotatebox{-90}                     
{\resizebox{10cm}{13cm}{\includegraphics{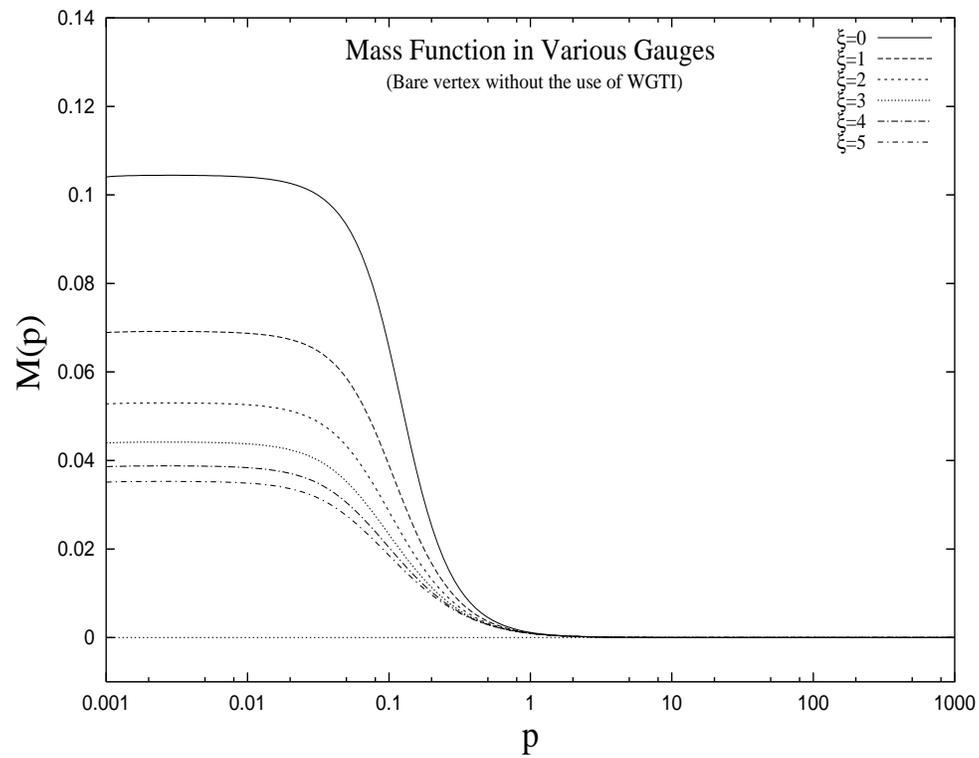}}}  
\vspace{5mm}                 
\caption{The mass function including the equation for the wavefunction 
renormalization.} 
\label{fnoWTI} 
\end{figure}          
\end{center}

\begin{center}
\begin{figure}[h]     
\rotatebox{-90}                     
{\resizebox{10cm}{13cm}{\includegraphics{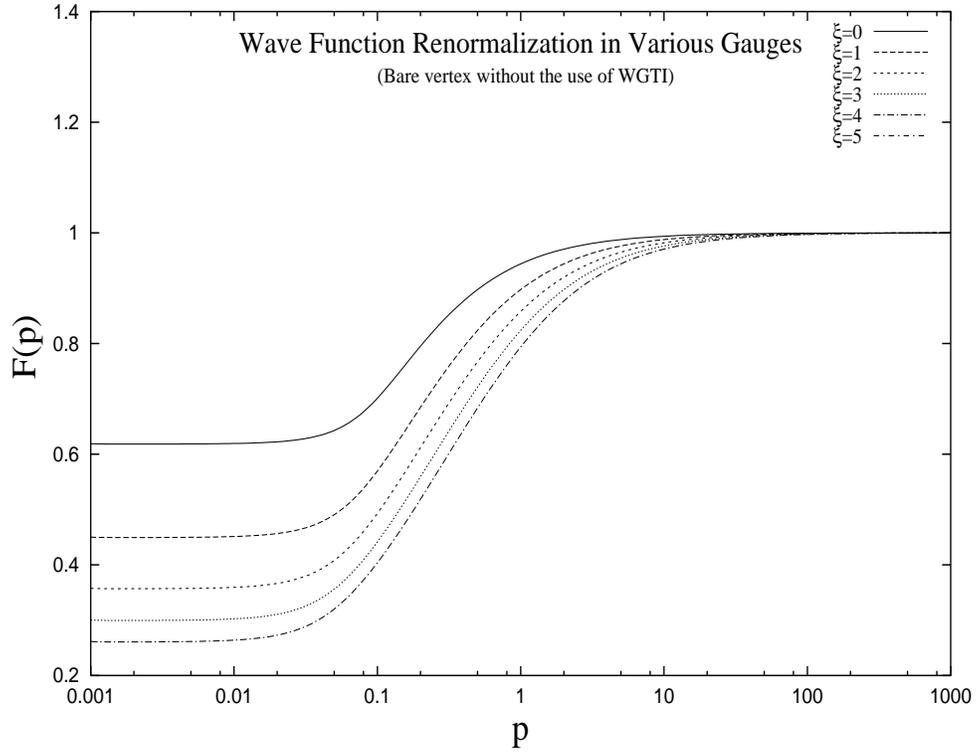}}}  
\vspace{5mm}                 
\caption{The wavefunction renormalization.} 
\label{fnoWTIwave} 
\end{figure}          
\end{center}

\begin{center}
\begin{figure}[h]     
\rotatebox{-90}                     
{\resizebox{10cm}{13cm}{\includegraphics{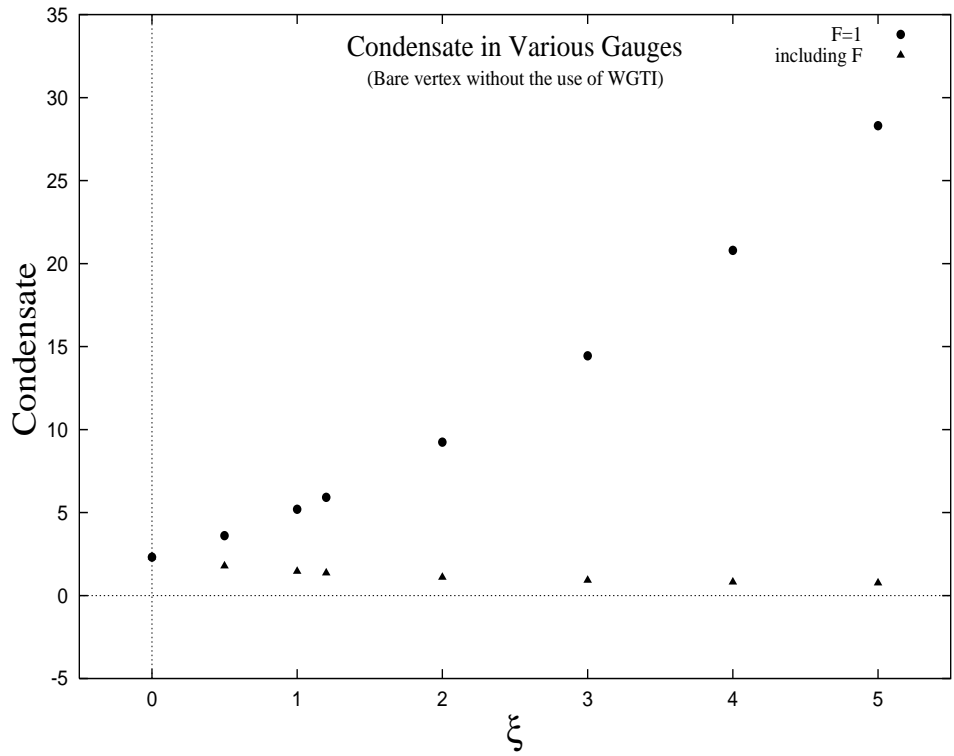}}}  
\vspace{5mm}                 
\caption{The condensate: a comparison between the cases with and without
the use of $F(p)=1$.} 
\label{fnoWTIcondcomp} 
\end{figure}          
\end{center}

\begin{center}
\begin{figure}[h]     
\rotatebox{-90}                     
{\resizebox{10cm}{13cm}{\includegraphics{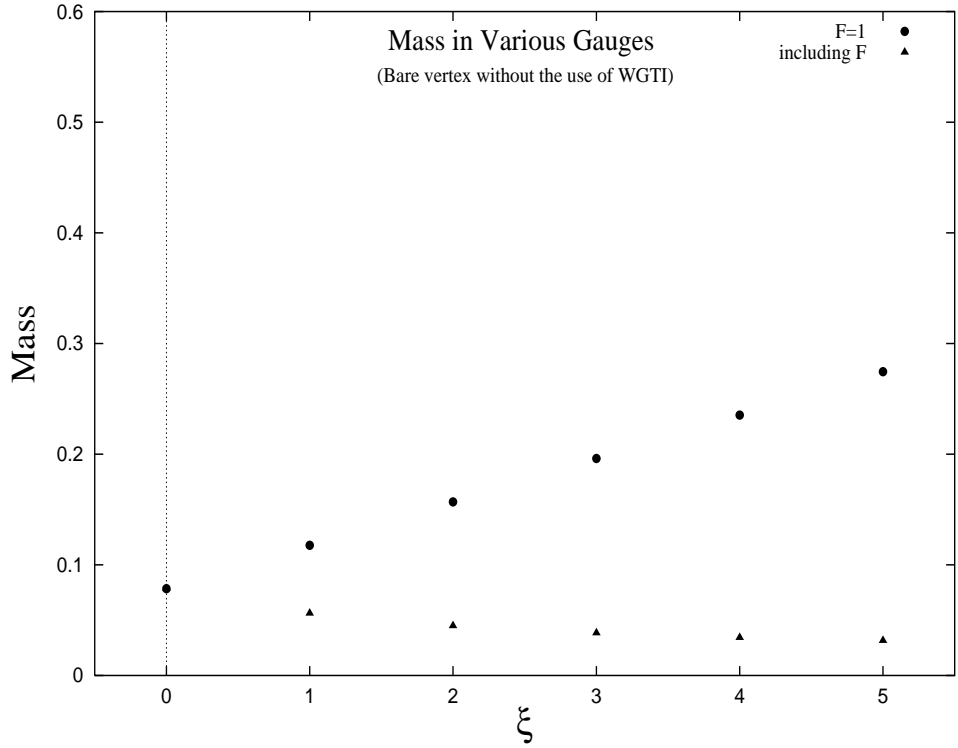}}}  
\vspace{5mm}                 
\caption{The mass: a comparison between the cases with and without
the use of $F(p)=1$.} 
\label{feucmass1} 
\end{figure}          
\end{center}

             Repeating the exercise by taking both the equations, namely 
Eqs.~(\ref{FNoWT}) and (\ref{MNoWT}), into account, we see
similar qualitative behaviour of the mass function. It is roughly a constant 
for low values of $p$ and falls as $1/p^2$ for large values of $p$, 
Fig.~(\ref{fnoWTI}).
The large $p$ behaviour is also evident from the entries in Table~(2).
As for the wavefunction renormalization, it also is constant for
small values of $p$. As $p$ becomes large it goes to $1$, 
Fig.~(\ref{fnoWTIwave}).
In Table~(2) we also give a comparison with the work of Burden and Roberts, 
\cite{BR1}. As mentioned earlier, they restrict themselves to the close 
vicinity of the Landau gauge, where our results are in excellent agreement.
We investigate the gauge dependence of the condensate as well as the 
mass far beyond the Landau gauge. A graphical description can be
found in Figs.~(\ref{fnoWTIcondcomp}) and (\ref{feucmass1}). The following
points are important to note~:
\begin{itemize}

\item The wavefunction renormalization plays an extremely important role in
restoring the gauge invariance of the chiral condensate as well as the
mass of the fermion. Although the qualitative behaviour of the
mass function in various regimes of momenta remains largely unchanged, 
whether or not we employ the approximation $F(p)=1$, quantitative dependence
of the physical observables mentioned above on the covariant gauge parameter
$\xi$ reduces a great deal by including the wavefunction renormalization.

\item As we move away from the Landau gauge towards large positive values
of $\xi$, the gauge dependence of the condensate as well as the mass
keeps diminishing, without resorting to any sophisticated {\em ansatze}
for the fermion-boson interaction. 

\end{itemize}

\vspace{5mm}
\hspace{2.5cm
\begin{tabular}{|crrr|}  \hline 
&     &     &   \\
\multicolumn{1}{|c}{$\xi$}  &
\multicolumn{1}{c}{$p$} &
\multicolumn{1}{c}{$\frac{4}{2+\xi}\,p^2\,M(p^2)$} &  
\multicolumn{1}{c|}{$\frac{4}{2+\xi}\,p^2\,M(p^2)$} \\
&     &     &   \\
&     & \small{BHR}    &  \small{BR} \\   \hline  \hline
&     &     &   \\
0.0   &  1000      &  2.31109   &  2.316 \\   
      &  642.233   &  2.31117   &   \\   
      &  316.228   &  2.31119   &   \\   
&     &     &   \\   \hline
&     &     &   \\
0.5   &  1000      &  1.77309   &  1.775 \\   
      &  642.233   &  1.77313   &   \\   
      &  316.228   &  1.77306   &   \\   
&     &     &   \\   \hline
&     &     &   \\
1.0   &  1000      &  1.44791   &  1.447 \\   
      &  642.233   &  1.44793   &   \\   
      &  316.228   &  1.44780   &   \\   
&     &     &   \\   \hline
&     &     &   \\
1.2   &  1000      &  1.35288   &  1.352 \\   
      &  642.233   &  1.35288   &   \\   
      &  316.228   &  1.35274   &   \\   
&     &     &   \\   \hline
&     &     &   \\
1.5   &  1000      &  1.23591   &   \\   
      &  642.233   &  1.23591   &   \\   
      &  316.228   &  1.23574   &   \\   
&     &     &   \\   \hline
&     &     &   \\
2.0   &  1000      &  1.09014   &   \\   
      &  642.233   &  1.09012   &   \\   
      &  316.228   &  1.08992   &   \\   
      &     &     &   \\   \hline   
\end{tabular}   
\begin{tabular}{|crrr|}  \hline 
&     &     &   \\
\multicolumn{1}{|c}{$\xi$}  &
\multicolumn{1}{c}{$p$} &
\multicolumn{1}{c}{ $\frac{4}{2+\xi}\,p^2 \,M(p^2)$ } &
\multicolumn{1}{c|}{$\frac{4}{2+\xi}\,p^2\,M(p^2)$} \\ 
&     &     &   \\
&     & \small{BHR}    &  \small{BR} \\   \hline  \hline
&     &     &   \\
2.5   &  1000      &  0.98597   &   \\   
      &  642.233   &  0.98594   &   \\   
      &  316.228   &  0.98571   &   \\ 
&     &     &   \\   \hline
&     &     &   \\   
3.0   &  1000      &  0.90941   &   \\   
      &  642.233   &  0.90938   &   \\   
      &  316.228   &  0.90912   &   \\   
&     &     &   \\   \hline
&     &     &   \\
3.5   &  1000      &  0.85201   &   \\   
      &  642.233   &  0.85197   &   \\   
      &  316.228   &  0.85169   &   \\   
&     &     &   \\   \hline
&     &     &   \\
4.0   &  1000      &  0.80839   &   \\   
      &  642.233   &  0.80833   &   \\   
      &  316.228   &  0.80803   &   \\   
&     &     &   \\   \hline
&     &     &   \\
4.5   &  1000      &  0.77497   &   \\   
      &  642.233   &  0.77491   &   \\   
      &  316.228   &  0.77458   &   \\   
&     &     &   \\   \hline
&     &     &   \\
5.0   &  1000      &  0.74933   &   \\   
      &  642.233   &  0.74927   &   \\   
      &  316.228   &  0.74891   &   \\   
&     &     &   \\   \hline
\end{tabular}}  \\ \\ \\
{\centerline {TABLE~2. The condensate in various gauges including wavefunction
renormalization}}
\vspace{5mm}

\section{EFFECT OF THE WARD-GREEN-TAKAHASHI IDENTITY}

\begin{center}
\begin{figure}[h]     
\rotatebox{-90}                     
{\resizebox{10cm}{13cm}{\includegraphics{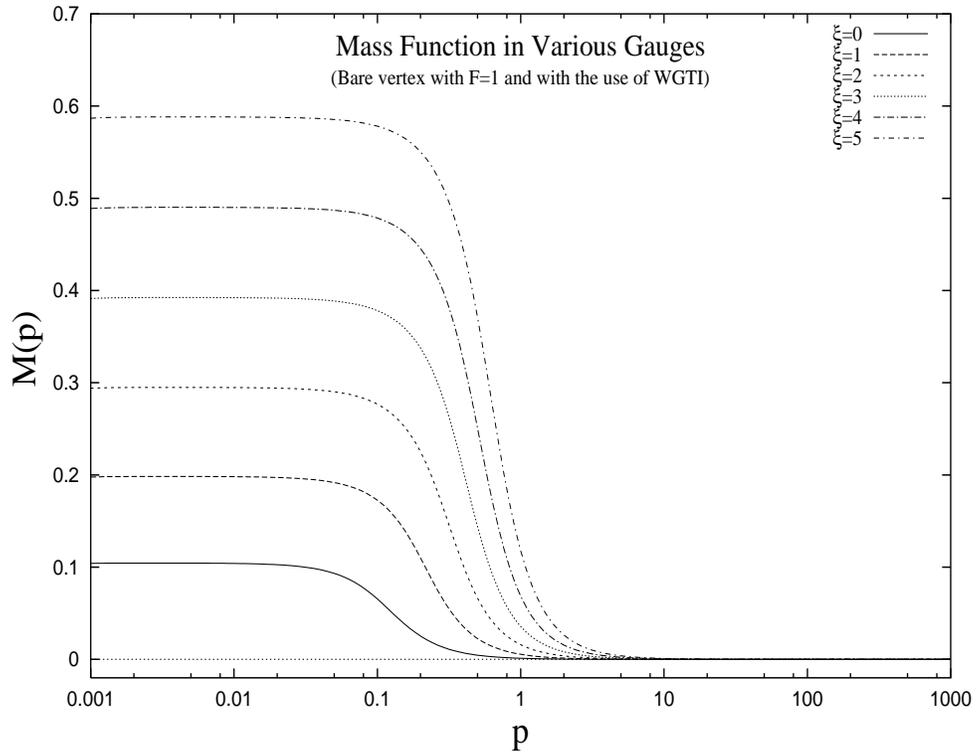}}}  
\vspace{5mm}                 
\caption{The mass function including the effect of the WGTI 
for $F(p)=1$.} 
\label{fWTIF1} 
\end{figure}          
\end{center}

\begin{center}
\begin{figure}[h]     
\rotatebox{-90}                     
{\resizebox{10cm}{13cm}{\includegraphics{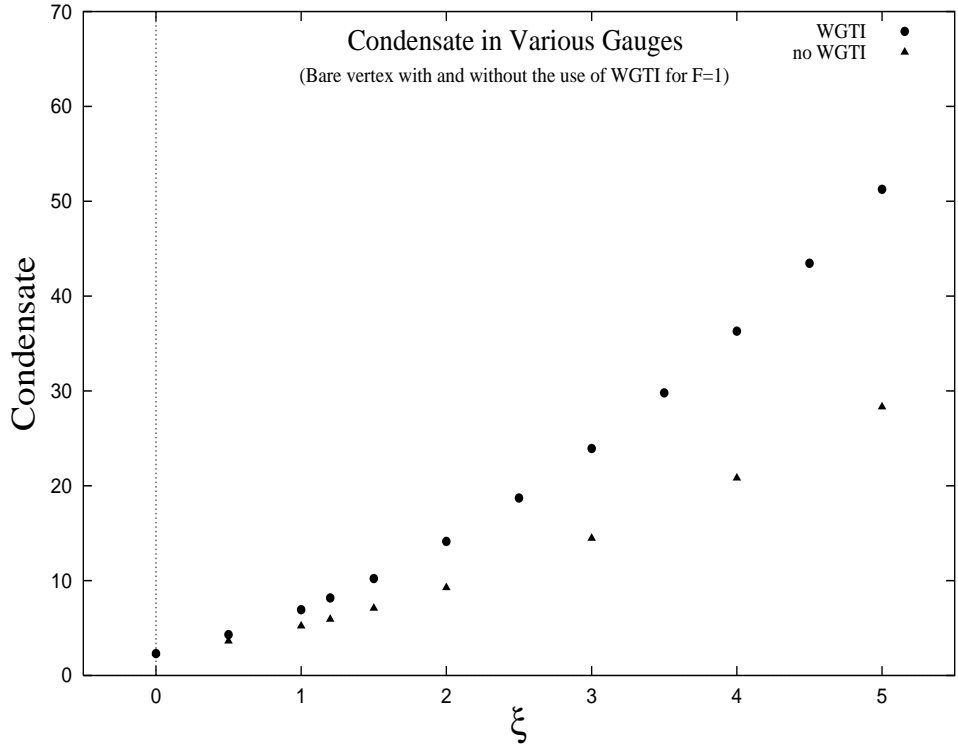}}}  
\vspace{5mm}                 
\caption{Effect of the WGTI on the gauge dependence of the 
condensate for $F(p)=1$.} 
\label{fcondF1WTIcompnoWTI}
\end{figure}          
\end{center}

\begin{center}
\begin{figure}[h]     
\rotatebox{-90}                     
{\resizebox{10cm}{13cm}{\includegraphics{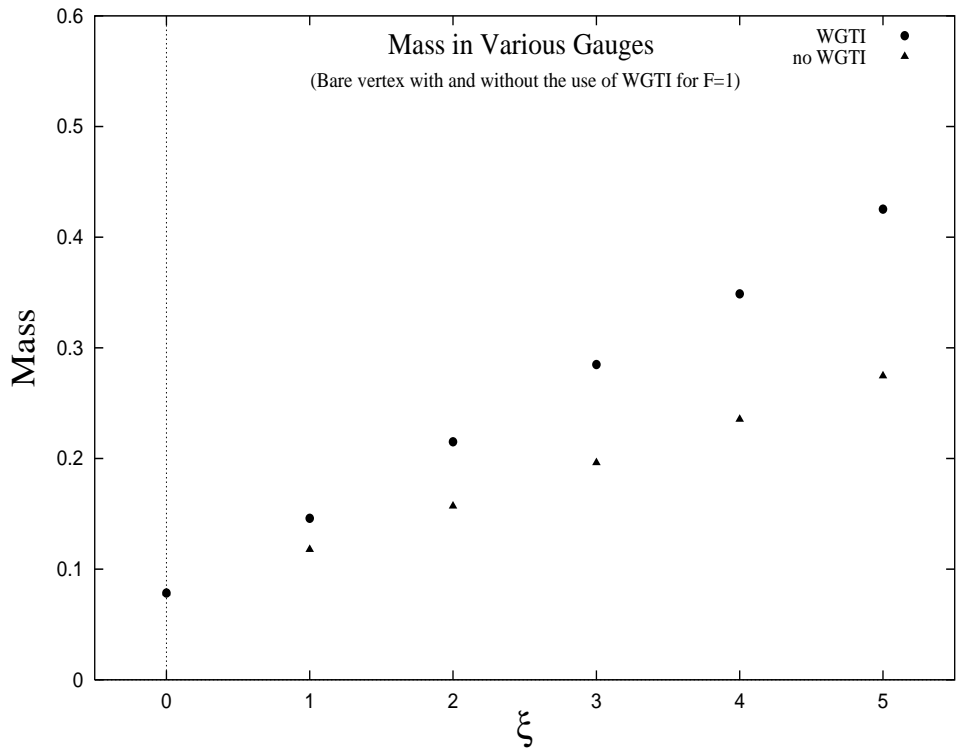}}}  
\vspace{5mm}                 
\caption{Effect of the WGTI on the gauge dependence of the 
 mass for $F(p)=1$.} 
\label{feucmass2}
\end{figure}          
\end{center}

The bare photon propagator which appears in Eq.~(\ref{prop}) can be split
up in longitudinal and transverse parts as follows~:
\be
\Delta^0_{\mu \nu}(q) =  \Delta^{0 T}_{\mu \nu}  -\xi \, 
\frac{ q_{\mu}q_{\nu} }{q^4} \:,
\ee
where $\Delta^{0 T}_{\mu \nu} = -g_{\mu \nu} / q^2 + 
q_{\mu}q_{\nu} / q^4$. Employing this decomposition, we can rewrite 
Eq.~(\ref{prop}) as~:
\be
S_F^{-1}(p) &=& S_F^{0 -1}(p) - 
ie^2\int \frac{d^3k}{(2\pi)^3} \Gamma^{\nu}(k,p)  S_F(k) 
\gamma^{\mu}\Delta^{0 T}_{\mu \nu}  
+ ie^2 \xi
\int \frac{d^3k}{(2\pi)^3} \Gamma^{\nu}(k,p) S_F(k) 
\gamma^{\mu} \frac{q_{\mu}q_{\nu}}{q^4} \label{propWTI} \;.
\ee
It is well known that the use of the WGTI, in the equivalent of the 
last term of Eq.~(\ref{propWTI}) in QED4, filters out a spurious term
which is an artifact of using the gauge dependent cut-off regulator.
Therefore, one is naturally motivated to use this decomposition in
dimensions other than four.
Multiplying Eq.~(\ref{propWTI}) by $1$ and $\not \! p$ respectively
and Wick-rotating to the Euclidean space, we obtain the following equations~:
\begin{eqnarray}
\frac{1}{F(p)}&=& 1 + \frac{\alpha}{2{\pi}^2p^2} \int d^3k
\, \frac{F(k)}{k^2+{\cal M}^2(k)} \, \frac{1}{q^4}
\left[\,2(q\cdot p)(q\cdot k) 
+ \, \frac{\xi}{F(p)}\,[p^2(q \cdot k)+{\cal M}(k){\cal M}(p)
(q\cdot p)]\;
\; \right]  \;,  \\
\frac{{\cal M}(p)}{F(p)}&=&  \frac{ \alpha}{2\pi^2} \int d^3k
\, \frac{F(k)}{k^2+{\cal M}^2(k)} \, \frac{1}{q^2} 
\left[\,2 {\cal M}(k)\, 
- \frac{\xi}{q^2}\, \frac{1}{F(p)}\,[{\cal M}(k)\, 
(p\cdot q) - {\cal M}(p)\,(k\cdot q)\,]
\; \right] \;. \label{Mangle}   
\label{Fangle}
\end{eqnarray}
On carrying out angular integration, 
\begin{eqnarray}
\frac{1}{F(p)}&=& 1 + \frac{\alpha \xi}{\pi p^2}\int_0^{\infty}dk
\, \frac{k^2  F(k)/F(p)}{k^2+{\cal M}^2(k)} 
\Bigg[
\frac{p^2}{k^2-p^2}+\frac{p}{2k} {\rm ln}\left| \frac{k+p}{k-p}  \right| 
 +{\cal M}(k){\cal M}(p)\left\{ 
\frac{1}{k^2-p^2}-\frac{1}{2kp} {\rm ln}\left| \frac{k+p}{k-p}  \right|
 \right\}
 \Bigg]  \;, \label{3 1/F}   \\
\frac{{\cal M}(p)}{F(p)}&=& \frac{\alpha}{\pi}\int_0^{\infty}dk
\, \frac{k^2 F(k)}{k^2+{\cal M}^2(k)} \Bigg[  
 \frac{2 {\cal M}(k)}{kp}{\rm ln}\left| 
\frac{k+p}{k-p} \, \right| 
- \frac{\xi}{F(p)} \left\{ \frac{{\cal M}(k) - 
{\cal M}(p)}{k^2-p^2} - 
\frac{{\cal M}(k)+{\cal M}(p)}{2kp}  {\rm ln}
\left|  \frac{k+p}{k-p} \right|  \right\}    
 \Bigg] \;.   \label{3 M/F}  
\end{eqnarray}
As the terms of the type $1/(k^2-p^2)$ are harder to deal with numerically,
we use the approximation $F(p)=1$ to analyze the effect of the 
WGTI. Under this simplification, we only have to solve
\begin{eqnarray}
{\cal M}(p)&=& \frac{\alpha}{\pi}\int_0^{\infty}dk
\, \frac{k^2 }{k^2+{\cal M}^2(k)} \Bigg[  
 \frac{2 {\cal M}(k)}{kp}{\rm ln}\left| 
\frac{k+p}{k-p} \, \right| 
- \xi \left\{ \frac{{\cal M}(k) - 
{\cal M}(p)}{k^2-p^2} - 
\frac{{\cal M}(k)+{\cal M}(p)}{2kp}  {\rm ln}
\left|  \frac{k+p}{k-p} \right|  \right\}    
 \Bigg] \;.   \label{3M}  
\end{eqnarray}
The mass function obtained on solving Eq.~(\ref{3M}) is depicted in
Fig.~(\ref{fWTIF1}), which, along with Table~(3), reveals that
its qualitative behaviour remains unchanged both for small and large values 
of $p$. In Figs.~(\ref{fcondF1WTIcompnoWTI}) and~(\ref{feucmass2}) , we 
compare the gauge dependence of the condensate and the mass with and 
without the usage of the WGTI. As in QED4, we find that the
gauge dependence of these quantities seems to increase by incorporating
the said identity.

\vspace{5mm}
\hspace{4.5cm
\begin{tabular}{|crr|}  \hline 
&     &    \\
\multicolumn{1}{|c}{$\xi$}  &
\multicolumn{1}{c}{$p$} &
\multicolumn{1}{c|}{$\frac{4}{2+\xi}\,p^2\,M(p^2)$}  \\
&     &    \\                                 \hline  \hline
&     &    \\
0.0 &  1000      &  2.31109     \\   
    &  642.233   &  2.31117     \\   
    &  316.228   &  2.31119     \\   
&     &    \\                        \hline
&     &    \\
0.5 &  1000      &  4.30196     \\   
    &  642.233   &  4.30221     \\   
    &  316.228   &  4.30248     \\   
&     &    \\                        \hline
&     &    \\
1.0 &  1000      &  6.93473     \\   
    &  642.233   &  6.93529     \\   
    &  316.228   &  6.93609     \\   
&     &    \\                        \hline
&     &    \\
1.2 &  1000      &  8.16826     \\   
    &  642.233   &  8.16900     \\   
    &  316.228   &  8.17011     \\   
&     &    \\                        \hline
&     &    \\
1.5 &  1000      &  10.2122     \\   
    &  642.233   &  10.2133     \\   
    &  316.228   &  10.2150     \\   
&     &    \\                        \hline
&     &    \\
2.0 &  1000      &  14.1356     \\   
    &  642.233   &  14.1375     \\   
    &  316.228   &  14.1406     \\   
  &     &    \\                               \hline   
\end{tabular}  
\begin{tabular}{|crr|}  \hline 
&     &    \\
\multicolumn{1}{|c}{$\xi$}  &
\multicolumn{1}{c}{$p$} &
\multicolumn{1}{c|}{$\frac{4}{2+\xi}\,p^2\,M(p^2)$}  \\
&     &    \\                                 \hline  \hline
&     &    \\
2.5 &  1000      &  18.7057     \\   
    &  642.233   &  18.7085     \\   
    &  316.228   &  18.7136     \\   
&     &    \\                        \hline
&     &    \\
3.0 &  1000      &  23.9226     \\   
    &  642.233   &  23.9268     \\   
    &  316.228   &  23.9345     \\   
&     &    \\                        \hline
&     &    \\
3.5 &  1000      &  29.7865     \\   
    &  642.233   &  29.7924     \\   
    &  316.228   &  29.8037     \\   
&     &    \\                        \hline
&     &    \\
4.0 &  1000      &  36.2975     \\   
    &  642.233   &  36.3056     \\   
    &  316.228   &  36.3212     \\   
&     &    \\                        \hline
&     &    \\
4.5 &  1000      &  43.4556     \\   
    &  642.233   &  43.4663     \\   
    &  316.228   &  43.4872     \\   
&     &    \\                        \hline
&     &    \\
5.0 &  1000      &  51.2608     \\   
    &  642.233   &  51.2746     \\   
    &  316.228   &  51.3020     \\   
  &     &    \\                               \hline   
\end{tabular}}  \\ \\ \\ 
{\centerline {TABLE~3. The condensate in various gauges for $F(p)=1$ making
use of WGTI}}
\vspace{5mm}

\section{DIMENSIONAL REGULARIZATION METHOD}
\begin{center}
\begin{figure}[h]     
\rotatebox{-90}                     
{\resizebox{10cm}{13cm}{\includegraphics{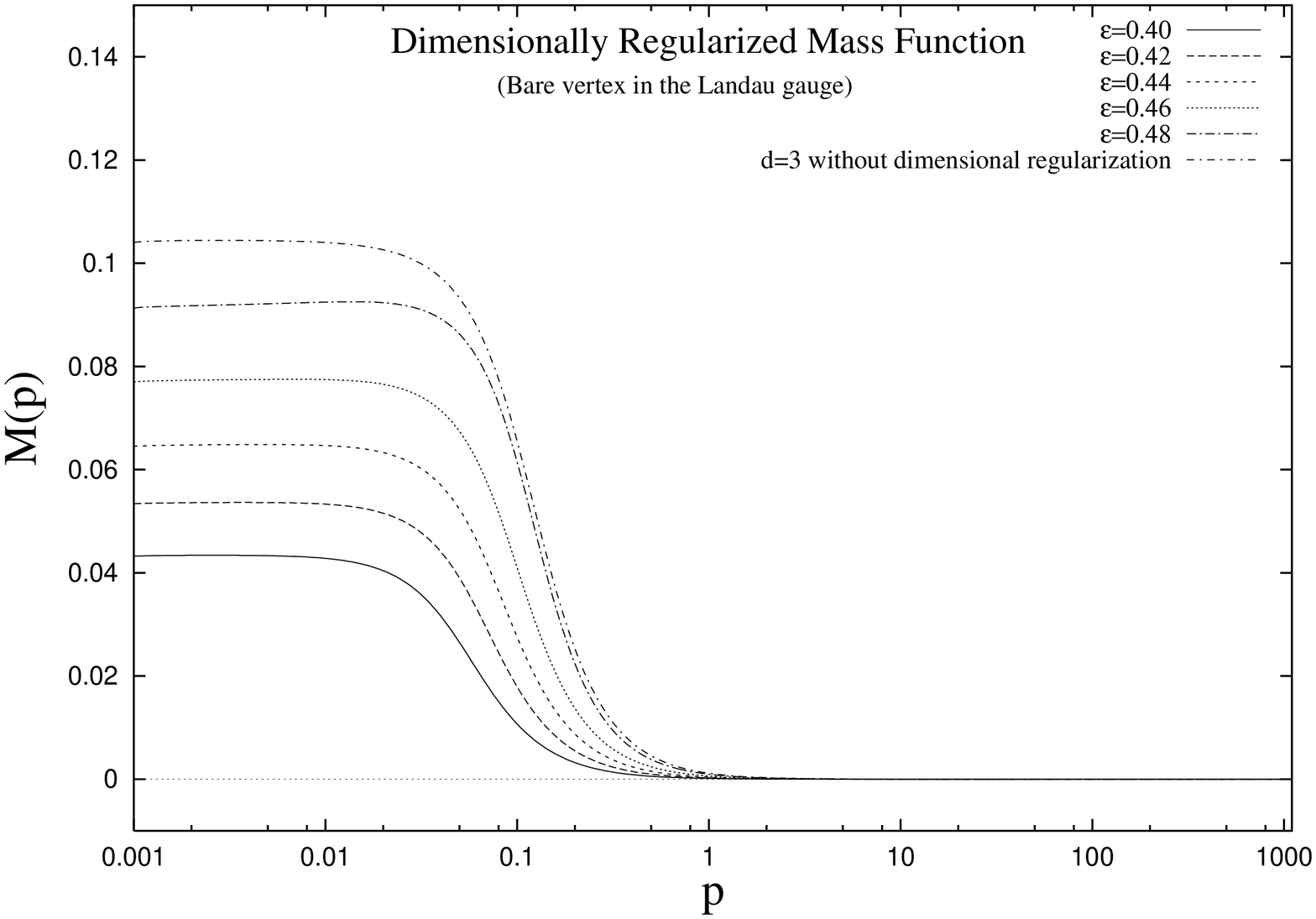}}}  
\vspace{5mm}                 
\caption{The mass function for various values of $\epsilon$. Note that 
$\epsilon=0.5$ would correspond to 3 dimensions. The result for 
$\epsilon=0.48$ 
could not be achieved up to the desired accuracy because of the ever increasing
computational time and memory involved.}
\label{fdimreg}
\end{figure}          
\end{center}

In this section we compare our numerical results with those obtained 
by employing the dimensional regularization scheme, \cite{GSSW,SSW,KSW}. For 
simplicity, we restrict ourselves only to the Landau gauge 
without incorporating the WGTI. In this case, the  
equation for the mass function acquires the following form in 
Euclidean space in arbitrary dimensions~:
\begin{equation}
{\cal M}(p)=4\pi\alpha (d-1)\int\frac{d^dk}{(2\pi)^d}
\frac{{\cal M}(k)}{k^2+{\cal M}^2(k)}\frac{1}{q^2},
\end{equation}
where $\alpha$ is a dimensionful coupling except in four dimensions. We define 
$d=4- 2 \epsilon$ and
\be
        \alpha &=& \alpha_d \; {\mu}^{2 \epsilon}  \;,
\ee
$\alpha_d$ being dimensionless. We now use the identity
$d^dk=k^{d-1} \,dk \, d \Omega_d$, where $d \Omega_d$ is the $d$-dimensional
solid angle defined as $d \Omega_d=\prod_{l=1}^{d-1} {\rm sin}^{d-1-l} \,
\theta_l \, d \theta_l $. The angle $\theta_{d-1}$ varies from 0 to
$2 \pi$, whereas all other angles vary from 0 to $\pi$. Choosing
$\theta_1$ to be the angle between $k$ and $p$, we can easily carry out
the remaining angular integrations to arrive at~:
\begin{eqnarray}
     {\cal M}(p) &=& \frac{2(d-1) \alpha}{(4 \pi)^{\frac{d-1}{2}} 
\Gamma(\frac{d-1}{2})}  \; \int_0^\infty dk^2 \; 
\frac{k^{d-2} \, {\cal M}(p)}{k^2 + {\cal M}^2(p)} \; 
\int_0^{\pi} \, d \theta_1 \; \frac{{\rm sin}^{d-2} \theta_1}{q^2} \;.
\end{eqnarray}
Using the standard formula, \cite{GR},
\begin{eqnarray*}
   \int_0^{\pi} \;dx \, 
\frac{{\rm sin}^{2 \sigma -1}x}{\left[1+2 a {\rm cos}x +a^2 \right]^{\lambda}}
&=& B(\sigma,1/2) \; F(\lambda,\lambda-\sigma+1/2,\mu+1/2;a^2) 
\hspace{15mm}  |a|<1 \;,
\end{eqnarray*}
integration over $\theta_1$ yields~:
\begin{eqnarray}
{\cal M}(p)&=& \frac{ (3-2 \epsilon) \alpha}{(4 \pi)^{1- \epsilon} 
\Gamma(2- \epsilon)}  \int_0^\infty dk^2
\frac{(k^2)^{1-\epsilon}{\cal M}(k)}{k^2+{\cal M}^2(k)}
\left[ \frac{1}{k^2}F\left( 1,\epsilon;2-\epsilon;\frac{p^2}{k^2}
\right) \theta (k^2-p^2)
+\frac{1}{p^2}F \left(1,\epsilon;2-\epsilon;\frac{k^2}{p^2}\right)
\theta (k^2-p^2)\right]. \nn \\
\end{eqnarray}
This equation was studied in detail in \cite{GSSW} in four dimensions,
taking $\epsilon$ to be a small positive number. The  factor 
$(k^2)^{- \epsilon}$ in the numerator regulates the otherwise
divergent behaviour of the integrand for large momenta. As noted
in \cite{GSSW}, the hypergeometric function does not play any 
role in regularization and hence can simply be replaced by 
$F(1,0;2,z)=1$. In case of three dimensions, the hypergeometric
function develops a pole for $k^2=p^2$, as is obvious from the
following identity~:
\begin{equation}
F(1,\epsilon;2-\epsilon;1)=\frac{1-\epsilon}{1-2\epsilon} \;. \label{pole}
\end{equation}
As was pointed out earlier, such terms are hard to deal with 
numerically. Due to increasing computational time 
and memory, we go only up to $\epsilon =0.48$, starting from 
$\epsilon=0.4$. To obtain satisfying results, we need to use increasingly
more points per decade as we approach closer to $\epsilon=0.5$. For instance,
 we use $100$ points per decade for $\epsilon =0.46$. The problems of
ever increasing computational time and memory limited us to use 
$140$ points per decade for the case of $\epsilon=0.48$. Despite this 
large number, we belive that the corresponding result falls short of the 
desired accuracy.
As a result, there is a slight rise at the end of the flat region of the
mass function, and the final descent begins rather late, Fig.~(\ref{fdimreg}).
 The problematic 
pole 
for $\epsilon=0.5$
in Eq.~(\ref{pole}) corresponds to the relatively well-controlled 
singularity in the following expression
\begin{eqnarray}
F\left( 1,\frac{1}{2};\frac{3}{2};z^2 \right) &=& \frac{1}{2z}
\ln{\frac{1+z}{1-z}}
\end{eqnarray}
for $z\rightarrow 1$. 
A comparison between the mass function obtained from techniques based
upon the dimensional regularization scheme and the one  computed in Section III
is also depicted in Fig.~(\ref{fdimreg}). Taking the numerical limitation 
for $\epsilon=0.48$ into account, we note that as 
$\epsilon$ approaches the value of $0.5$, we get 
closer and closer to the result obtained in Section III, where we
work in 3-dimensions to start with.

\section{Conclusions}

The sources of gauge non-invariance in the study of SDE in QED can arise 
from (i) the use of the approximation $F(p)=1$, 
(ii) the violation of WGTI and the LKF transformations,
(iii) the flawed {\em ansatz} for the vertex and (iv) the inadequate
choice of the regulator. In case of 3 dimensions, QED becomes 
neater because of the absence of ultraviolet divergences. The
regulator dependence does not pose a threat any longer. In this paper,
we have studied the quantitative effect of the approximation 
$F(p)=1$ and the violation of the WGTI on the gauge dependence
of the mass and the condensate in QED3 for a wide range of values of the
covariant gauge parameter $\xi$. Lifting the approximation $F(p)=1$ seems to 
reinstate gauge invariance to an impressive extent. Partial use of the
WGTI seems to increase the gauge dependence, an effect noted also in 4 
dimensions in \cite{GSSW}. As an advantage of using a broad range of values
of $\xi$, we find that when they are large,
the gauge dependence of the said quantities diminishes without resorting
to improved {\em ansatz} for the vertex. Currently, work is underway to 
explore the role played by various elaborate choices of the vertex in this 
context.

\newpage

\newpage

\newpage

\newpage

\newpage

\newpage

\newpage


\begin{thebibliography}{55}
\bibitem{LK1} L.D. Landau and I.M. Khalatnikov, Zh. Eksp. Teor. Fiz. {\bf 29} 
89 (1956).
\bibitem{LK2} L.D. Landau and I.M. Khalatnikov, Sov. Phys. JETP {\bf 2} 69
(1956). 
\bibitem{F1} E.S. Fradkin, Sov. Phys. JETP {\bf 2} 361 (1956).
\bibitem{JZ1} K. Johnson and B. Zumino, Phys. Rev. Lett. {\bf 3} 351 (1959).
\bibitem{Z1} B. Zumino, J. Math. Phys. {\bf 1} 1 (1960).
\bibitem{W1} J.C. Ward, Phys. Rev. {\bf 78} (1950).
\bibitem{G1} H.S. Green, Proc. Phys. Soc. (London) {\bf A66} 873 (1953).
\bibitem{T1} Y. Takahashi, Nuovo Cimento {\bf 6} 371 (1957).
\bibitem{Salam1} A. Salam, Phys. Rev. {\bf 130} 1287 (1963).
\bibitem{SD1} A. Salam and R. Delbourgo, Phys. Rev. {\bf 135} 1398 (1964).
\bibitem{S1} J. Strathdee, Phys. Rev. {\bf 135} 1428 (1964).
\bibitem{DW1} R. Delbourgo and P. West, J. Phys. {\bf A10} 1049 (1977).
\bibitem{DW2} R. Delbourgo and P. West, Phys. Lett. {\bf B72} 96 (1977).
\bibitem{D1} R. Delbourgo, Nuovo Cimento {\bf A49} 484 (1979).
\bibitem{D2} R. Delbourgo, Austral. J. Phys. {\bf 52} 681 (1999).
\bibitem{CP1} D.C. Curtis and M.R. Pennington, Phys. Rev. {\bf D42} 4165 
(1990).
\bibitem{CP2} D.C. Curtis and M.R. Pennington, Phys. Rev. {\bf D44} 536 
(1991).
\bibitem{CP3} D.C. Curtis and M.R. Pennington, Phys. Rev. {\bf D48} 4933 
(1993).
\bibitem{ABGPR1} D. Atkinson, J.C.R. Bloch, V.P. Gusynin, M.R. Pennington
and M. Reenders, Phys. Lett. {\bf B329} 117 (1994).
\bibitem{AGM1} D. Atkinson, V.P. Gusynin and P. Maris, Phys. Lett. {\bf B303}
157 (1993).
\bibitem{BP1} A. Bashir and M.R. Pennington, Phys. Rev. {\bf D50} 7679 (1994).
\bibitem{BP2} A. Bashir and M.R. Pennington, Phys. Rev. {\bf D53} 4694 (1996).
\bibitem{GSSW} V.P. Gusynin, A.W. Schreiber, T. Sizer and A.G. Williams,
Phys. Rev. {\bf D60} 065007 (1999).
\bibitem{SSW} A.W. Schreiber, T. Sizer and A.G. Williams, Phys. Rev. 
{\bf D58} 125014 (1998).
\bibitem{KSW} A. K{\i}z{\i}lers\"{u}, A.W. Schreiber and A.G. Williams, 
Phys. Lett {\bf B499} 261 (2001).
\bibitem{BR1} C.J. Burden and C.D. Roberts, Phys. Rev. {\bf D44} 540 (1991).
\bibitem{BKP1} A. Bashir, A. K{\i}z{\i}lers\"{u} and M.R. Pennington, 
Phys. Rev. {\bf D57} 1242 (1998).
\bibitem{AB1} A. Bashir, Phys. Lett. {\bf B491} 280 (2000).
\bibitem{GR} I.S. Gradshteyn and I.M. Ryzhik, {\em Table of Integrals,
Series and Products}, sixth edition (Academic Press, USA), 2000.
\end{thebibliography}
\end{document}